# A transportable $^{40}$Ca$^+$ single-ion clock with 7.7×10$^{-17}$ systematic uncertainty

Jian Cao[1,2,], Ping Zhang[1,2,3,4], Junjuan Shang[1,2,3], Kaifeng Cui[1,2,3], Jinbo Yuan[1,2,3], Sijia Chao[1,2,3],

Shaomao Wang[1,2,3], Hualin Shu[1,2], Xueren Huang[1,2,†]

[1]*State Key Laboratory of Magnetic Resonance and Atomic and Molecular Physics, Wuhan Institute of Physics and Mathematics,*
*Chinese Academy of Sciences, Wuhan 430071, China*

[2]*Key Laboratory of Atomic Frequency Standards, Wuhan Institute of Physics and Mathematics,*
*Chinese Academy of Sciences, Wuhan 430071, China*

[3]*University of Chinese Academy of Sciences, Beijing 100080, China*

[4]*School of Physics and Electrical Engineering, Taizhou University, Taizhou 318000, China*

† hxueren@wipm.ac.cn

**Abstract**: A transportable optical clock refer to the 4s$^2$S$_{1/2}$-3d$^2$D$_{5/2}$ electric quadrupole transition at 729 nm of single $^{40}$Ca$^+$ trapped in mini Paul trap has been developed. The physical system of $^{40}$Ca$^+$ optical clock is re-engineered from a bulky and complex setup to an integration of two subsystems: a compact single ion unit including ion trapping and detection modules, and a compact laser unit including laser sources, beam distributor and frequency reference modules. Apart from the electronics, the whole equipment has been constructed within a volume of 0.54 m$^3$. The systematic fractional uncertainty has been evaluated to be 7.7×10$^{-17}$, and the Allan deviation fits to be $2.3 \times 10^{-14} / \sqrt{\tau}$ by clock self-comparison with a probe pulse time 20 ms.

**Keywords**: Optical clock; Ion trap; Transportable.

## 1. Introduction

Due to the impressive progress in recent years, optical clocks are deemed for the redefinition of SI second in future as they have surpassed the performance of Cs primary clocks in both accuracy and stability[1-5]. At the same time, these ultra-precise optical clocks offer new possibilities for high precision tests of fundamental physics[6-9], geophysics[10; 11], improved timekeeping and satellites navigation[12; 13]. On the other hand, a more accurate and direct mean of optical frequency comparison is required for today's best clocks over quite long distance. Since the uncertainty of satellite links are limited and dedicated equipments should be employed, and optical fiber links are not feasible enough especially for intercontinental comparison[14; 15]. For these reasons, it's quite necessary to develop the applicable optical clocks on ground and even in the space.

It's feasible to develop a transportable optical clock in the first stage. Essential characteristics for a transportable optical clock are: compact size and proper mass, ensuring at the same time a high performance and reliability, but also fully automatic operation. Undertaking a necessary step towards optical clocks in space, the "Space Optical Clock" program of European Space Agency (ESA) aims at accurate transportable $^{87}$Sr and $^{171}$Yb optical lattice clock demonstrators with an expected performance of fractional uncertainty below 5×10$^{-17}$ and fractional instability below $1 \times 10^{-15} / \sqrt{\tau}$ [16]. At present, a transportable $^{87}$Sr optical lattice clock (excluding electronics) fits within a volume of <2 m$^3$ with uncertainty of 7×10$^{-15}$ have been developed in LENS of Italy[17]. Other similar prototypes are under development in PTB, HHUD and NPL[18-20].

In addition to neutral atoms, a trapped $^{40}$Ca$^+$ single-ion is another promising option in the programs of transportable optical clocks. These kinds of optical clock have their own advantages on the relatively simple scheme of clock system, especially the simplicity on all semiconductor lasers that allowing for developing compact, robust and low-cost applicable clock. Moreover, an excellent performance of frequency uncertainty and instability make them quite attractive and cost-effective[21]. In this article, we present the realization of a transportable optical clock refer to the $4s^2S_{1/2}$-$3d^2D_{5/2}$ electric quadrupole transition at 729 nm of $^{40}Ca^+$ single-ion trapped in miniature Paul trap. Apart from the electronics, the whole equipment has been constructed within a volume of 0.54 m$^3$. The overall systematic fractional uncertainty has been evaluated to be 7.7×10$^{-17}$, and the Allan deviation fits to be $2.3\times10^{-14}/\sqrt{\tau}$ by clock self-comparison with a probe pulse time 20 ms.

2. **Clock design**

The design of the first-generation transportable $^{40}Ca^+$ single-ion optical clock , which is fully operational in our laboratory, is shown in Fig. 1. The physical system of this apparatus is re-engineered from a bulky and complex setup to an integration of two subsystems: a compact single ion subsystem including ion trapping and detection modules, and a compact laser subsystem including laser sources, beam distributor and frequency stabilization modules. All the connections between each modules are provided in single-mode polarization-maintaining fibers. This kind of modular design ensures better stability and reliability essential for long-term operation of transportable clock. Moreover, the development in the field of commercial optoelectronic subcomponents allows for independent testing and maintenance of each modules, such as a simple replacement or promotion of components.

The ion system consists of a compact vacuum chamber with the trap region takes the form of a cuboid (size:7×7×7 cm$^3$) which is kept to less than 5×10$^{-8}$ Pa by a 10 L/s ion pump. A miniature ring-endcap ion trap optimized by finite element analysis method, with a center-to-ring electrode distance of $r_0 \approx 0.75$ mm and an endcap-to-center distance of $z_0 \approx 0.75$ mm[22], is driven by a helical resonator at $\Omega_{rf} \approx 2\pi \times 24.54(10)$ MHz at which the rf-induced 2$^{nd}$-order Doppler frequency shifts and Stark shift cancel each other approximately[23; 24]. The background magnetic field is reduced by two layers of magnetic shielding to $B \approx 2.2$ μT. Laser induced fluorescence emitted by the $^{40}Ca^+$ single-ion at 397 nm is collected by a custom-made objective which is fixed to the vacuum window to ensure the stability of the module during transporting. All the photons collected by the objective are send to a photomultiplier (PMT), and both a narrow bandpass filter and an aperture stop are employed to filt the background stray light. The efficiency of this fluorescence detecting module is about 0.002, considering the collection of the solid angle, the transmittance of optical elements and the quantum efficiency of the PMT.

The design of the compact laser system is modular and mainly consists of the following modules: laser source, laser beam distributor and laser frequency reference. The laser source box contains all of the 5 external cavity diode lasers (ECDLs) employed for $^{40}Ca^+$ single-ion optical clock, and the wavelength are 423 nm (photo-ionization of calcium atoms), 397 nm (Doppler cooling and fluorescence detection), 866 nm (repumping for keeping the cycles of cooling), 729 nm (clock transition inquiring) and 854 nm (quenching from the metastable state after clock inquiring). Through the laser beam distributor, each output of these ECDLs is splitted into a few parts and then are coupled to wavemeter, the ion system and modules of laser frequency reference, respectively.

The module of laser frequency reference contains two Fabry-Pérot cavities. One is a 3-in-1 ultralow-expansion (ULE) glass cavity used for three lasers (397/866/854 nm, fineness ∼ 100) simultaneously by three pairs of fused-silica mirrors for each wavelength. Due to the cavity is made of ULE glass and further sealed in a thermal shield and vacuum chamber which are encapsulated in a box (size:52×46×34 cm$^3$), the frequency fluctuation of these lasers is about 1 MHz/day which is acceptable for daily operation. The other one is an all ULE glass cavity including mirrors (fineness ～300,000) used for 729 nm clock laser. By means of finite element analysis, the shape of the cavity and the supporting way are optimized and the setup is placed on a small active vibration isolated platform to get a vibration-insensitive performance. A set of thermal shield and a temperature controlled vacuum chamber as similar as the one used for 3-in-1 cavity are also employed and all of these devices including compact designed optical components are encapsulated in a box (size: 52×46×45 cm$^3$) to get a thermal-insensitive performance. After the free running 729 nm laser is locked to this ultra-stable cavity by Pound-Drever-Hall (PDH) method[25], the linewidth is measured to be 1Hz at 1s and frequency instability is measured to be less than 2×10$^{-15}$ at 1-100 s after removing the linear drift which is always at the level of 0.05 Hz/s.

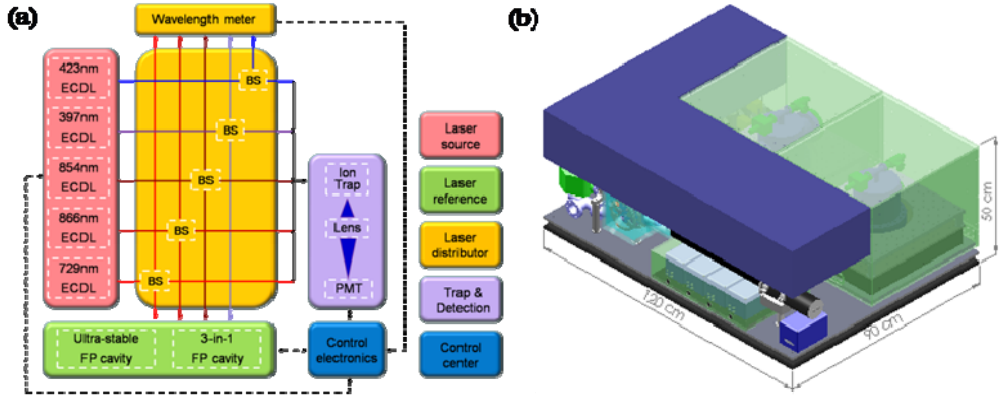

FIG. 1. Overview of the transportable $^{40}Ca^+$ single-ion clock. (a) Schematic diagram of clock system. (b) Design effect drawing of clock system. The blue "L" module is control electronics and the part that is not fully exposed under the electronics is the trap and detection module and laser source module. The rest of the space is mainly occupied by the laser reference module, which is represented by the green boxes.

### 3. Clock running and evaluation of its performance

A single $^{40}Ca^+$ ion is loaded into the trap by photoionizing neutral calcium atoms with a 423 nm diode laser and a 370 nm LED effectively. Typically, ~10 μW of the 397 nm laser power is focused on the ion with a beam waist of ~50 um, ~350 μW/350 μW/10 nW of power with ~100 μm beam size for the 866/854/729 nm laser, respectively. After micromotion minimization, the single ion can be trapped about one day continuously with laser Doppler cooling. An on-resonance fluorescence count rate of 40,000 /s is obtained for a single ion. For a typical background count rate of 2,000 /s this results in a signal-to-background ratio of 20. The secular motion frequencies of trapped ions are $\omega_{(r,z)} \approx 2\pi \times (2.1, 4.2)$ MHz when the forward power to the helical resonator is 5 W routinely, which are measured by 'tickling' the ions with low-power rf field applied to the compensation electrodes and endcap electrodes, respectively.

The 729 nm laser is locked to the TEM00 mode of the ultra-stable cavity which is the closest to the clock transition frequency. This frequency difference is covered with a double-passed acousto-optic modulator (AOM) driven by a signal generator referenced to an active hydrogen maser. This computer controlled signal generator updates the probe laser frequency every 40 cycles of pulses for spectroscopic measurements and locking to the transition line center by the "four points locking scheme"[26]. At the same time, the double-passed AOM also serves as a shutter for the probe laser beam. For cancellation of the linear Zeeman shift, the quadrupole shift and the tensor part of the Stark shifts, the laser is locked to the three inner pairs of Zeeman components with magnetic sublevels $m_{j'} = \pm 1/2, \pm 3/2, \pm 5/2$ in the upper $3d^2D_{5/2}$ state[27], sequentially in each cycle during the lock runs [Fig. 2(a)]. Each of the 6 components is measured independently and the resulting line center is combined by averaging the center frequencies of the 3 pair of components.

In order to prevent large ac-Stark shift of the 729 nm clock transition caused by the other 397/866/854 nm lasers, a pulse light sequence is introduced to observe the spectra. In this experiment, the cooling pulse, which includes the 397/866/854 nm radiations, is 15 ms. The 729 nm probe pulse is various according to the experimental requirements, and it induces a Fourier limited linewidth of 22 Hz with pulse time 40 ms [Fig. 2(b)], while all the other radiations are blocked by mechanical shutters. Buffer times of 3 ms are used before and after the probe pulse to again prevent ac-Stark shifts as the actuation time of each shutter is ~2 ms. After the clock probing pulse, the state of the ion is interrogated within 1ms using 397 nm and 866 nm lasers.

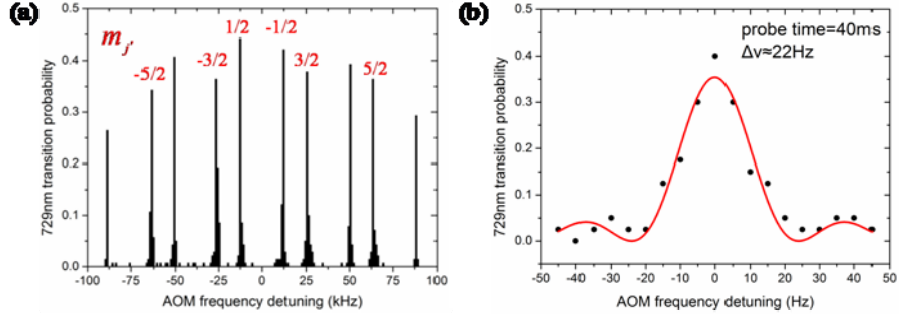

FIG. 2. Zeeman spectra of the $4s^2S_{1/2}$-$3d^2D_{5/2}$ clock transition. (a) Ten components of the Zeeman profile with the whole separation of ~180 kHz in the magnetic field of ~2.2 μT. The values of $m_{j'}$ of magnetic sublevels in the upper $3d^2D_{5/2}$ state selected for clock operation are shown in red marks. (b) Rabbi excitation spectrum of the $m_{j'}=1/2$ component with 729 nm pulse time 40 ms shows the spectrum resolution at 22 Hz level.

The thermal motion of the ion in the rf trap with secular motion frequency $\omega_{r,z}$ causes Doppler frequency shift. For our ring-endcap trap, the ion temperature is estimated from the intensity of the secular sidebands relative to the carrier[28]. These ratios normally are 0.02~0.05 for the radial sideband and 0.005~0.012 for the axial sideband, yielding a mean temperature of 6(1) mK which is nearly an order higher than the Doppler cooling limit. But thanks to the high secular motion frequencies, the single ion is laser-cooled to the Lamb-Dicke regime as the Lamb-Dicke approximation is satisfied well and therefore the first-order Doppler shift should not be considered. With the ion temperature estimated, the related second-order Doppler shift is calculated to be -8.6(1.4) mHz[29]. On the other side, thermal secular motion can push the ion into nonzero mean-square electric fields from the trap, which introduce a Stark shift. This shift can also be calculated from the ion temperature above for the selected three pairs of transition and the average shift is 16.7(2.8) mHz[29].

As similar as the thermal motion, the excess micromotion can also produce a second-order Doppler shift and a Stark shift. Since it's always one of the greatest contributions of systematic uncertainty for optical clock based on the ion in the rf trap, reduction of this micromotion to extremely low level is particularly important. Adjustment of the DC voltages on two compensation electrodes in the radial plane and one of the endcap electrodes while observing the trap region with both the EMCCD and PMT as the ion alternates between weak and tight trap confinement (rf power: 2-5 W) allows observation of possible changes of ion position and fluorescence counts by any residual trap asymmetry and stray fields[23]. This method rapidly converges the system to a low state of micromotion which is parallel to the EMCCD sensor and thus the upper limit can be estimated[21]. Ion displacement perpendicular to the EMCCD sensor will cause micromotion on the *k*-vector direction of the 729 nm laser beam, which can be estimated by measuring the ratio of micromotion sideband relative to carrier intensity. This ratio normally is 0.001~0.008 and the related second-order Doppler shift is calculated to be $\Delta\nu_{\mu,D2} = -18(18)$ mHz[23]. However, considering the Stark shift due to excess micromotion, the combined shifts can be shown as

$$\Delta\nu_\mu = \Delta\nu_{\mu,D2}[1 + \Delta\alpha_0 (mc\Omega_{rf})^2 / \hbar\omega_0 e^2], \quad (1)$$

where $\Delta\alpha_0$ is the differential static scalar polarizability of clock transition, $m$ the mass of ion, $\omega_0$ the clock transition angular frequency, $c$ the speed of light, $\hbar$ the Plank's constant, and $e$ the charge of an electron. For ions with a negative value of $\Delta\alpha_0$, such as $^{40}$Ca$^+$, the term in square brackets vanishes when the rf frequency takes the value of $\Omega_0 = e\sqrt{-\hbar\omega_0/\Delta\alpha_0}/mc$. For our trap, which currently works at $\Omega_{rf} \approx 2\pi \times 24.54(10)$ MHz, the term in the square brackets is 0.018(16) and therefore the micromotion shifts are suppressed by a factor of 63[27]. When applied to the value of second-order Doppler shift, the combined shifts are canceled down to a level of ±0.29(29) mHz. The estimated total micromotion shifts is thus -0.33(58) mHz and the tiny residual shift is produced from the little lower trap frequency.

The electric quadrupole shift caused by the interaction between the quadrupole moment and electric field gradient is always a great source of systematic shift. This shift of a magnetic sublevel $m_{j'}$ is given by[30]

$$\Delta f_{EQ} = \nu_{EQ} (3\cos^2\theta - 1)(m_{j'}^2 - j'(j'+1)/3)/4, \quad (2)$$

where $\nu_{EQ}$ is a characteristic frequency proportional to the electric quadrupole moment and the electric field gradient, and $\theta$ is the angle between the applied magnetic field and the electric field gradient. For the selected three pairs of Zeeman components with $m_{j'} = \pm 1/2, \pm 3/2, \pm 5/2$, there is

$$\sum_{m_{j'}=-j'}^{j'} (m_{j'}^2 - j'(j'+1)/3) = 0. \quad (3)$$

It means the net electric quadrupole shift can be canceled to an extremely high level naturally. Measurements of the selected three pairs of Zeeman components in Fig. 3(a) provides a diagnostic of the lock performance as the three shifts of $\Delta f_{EQ}$ must follow a linear dependence on $m_{j'}^2$. Histogram of measurements of electric quadrupole shift magnitude $A_{EQ} = \nu_{EQ}(3\cos^2\theta - 1)/4$ shown in Fig. 3(b) is also in reasonable agreement with Gaussian distribution. The uncertainty of $A_{EQ}$ is employed to evaluate the electric quadrupole shift by a worst-case scenario and thus the total frequency shift is estimated to be 0(31) mHz[27].

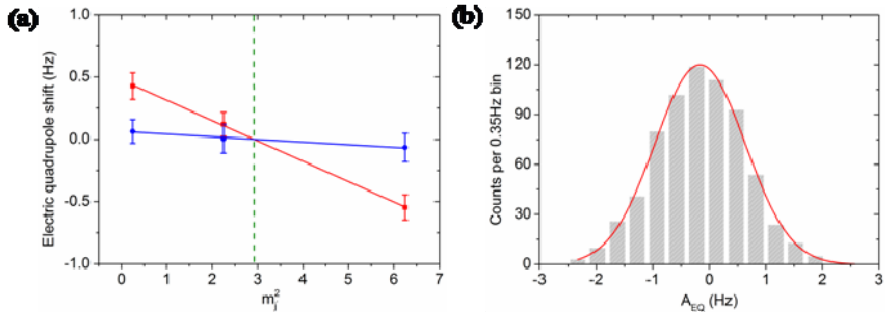

FIG. 3. Measurement of the electric quadrupole shift. (a) Measurement of three pairs of components provides a diagnostic of the lock performance as the three center frequencies must follow a linear dependence on $\Delta m_{j'}^2$. For $^{40}$Ca$^+$, the electric quadrupole shift is canceled at the intercept of $\Delta m_{j'}^2 = 35/12$. The red and blue dots are corresponding to the measurements at different electric field gradient, and the red ones are routinely observed in our experiments. (b) A histogram of the measurements of $A_{EQ} = \nu_{EQ}(3\cos^2\theta - 1)/4$. $A_{EQ}$ is determined experimentally using the frequency difference between shifted pairs: $A_{EQ} = [\Delta f_{EQ}(m_{j'} = 5/2) - \Delta f_{EQ}(m_{j'} = 1/2)]/6$.

The largest frequency shift in our transportable clock is the ac Stark shift induced by the blackbody radiation (BBR) emitted by the ion's environment. Both the temperature difference of the trap relative to the vacuum chamber and the temperature fluctuation of the vacuum chamber are evaluated for our system. A main complication in this evaluation is that the strong rf trap field will heat the dielectrics which are necessarily incorporated into the structure of an ion trap to electrically insulate various components. So the finite element models (FEM) is employed to calculate the temperature rise of any part of the trap structure and the resulting steady-state temperature distribution due to thermal conduction and radiation[31]. Then these results of simulation can be compared with experimental measurements by thermistor[3] and infrared camera[32] at critical test points in a dummy trap in vacuum chamber, which almost identical to the system used for the transportable clock. This comparison is essential for validation of the model to account for fabrication tolerances, and as some material and surface characterizations which are not clearly known. As indicated in Fig. 4, the temperature rise of the hottest part of the trap setup is 13.19 K under standard clock operation conditions ($\Omega_{rf} \approx 2\pi \times 24.54(10)$ MHz, $V_{0P} \approx 700(50)$ V), and the effective temperature rise at the position of ion is 1.72 K, with a deviation of 0.44 K mostly attribute to the uncertainty of trap electrodes emissivity. On the other side, the temperature at the position of ion is also evaluated to be 292.20 K, with a deviation of 0.12 K due to the temperature fluctuation of the vacuum chamber and electrodes of feedthrough. As a combination, the effective temperature at the position of ion is 293.92(0.46) K, and it results in a BBR frequency shift of 351.2(5.3) mHz[24; 33].

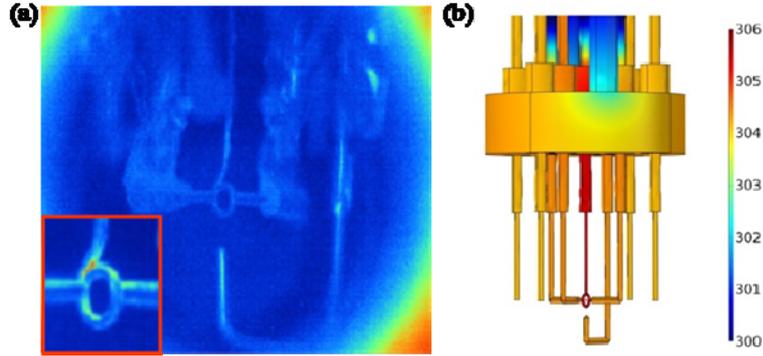

FIG. 4. Evaluation of the temperature distribution of trap. (a) Infrared camera images of the dummy trap structure. The large photograph is taken when rf power is off, and the inset is an amplification of the ring and endcap electrodes when rf power is on. The temperature rise from this comparision will be corrected with a calibrated PT1000. (b) FEM simulation of the temperature distribution of the operational trap for an applied rf potential of 700 V amplitude at 24.54 MHz. A maximum temperature rise of 13.19 K is found in the ring electrode.

The linear Zeeman shift is effectively cancelled out in our experiment by probing on symmetric components. However, the drift of the magnetic field can produce a residual linear Zeeman shift of 29.9 Hz/nT as there's a time interval of ~2.0 s between each probe point in the clock running. This drift of magnetic field is less than $3\times10^{-5}$ nT/s in a month measurements, and thus the residual linear Zeeman shift is estimated to be 0.0(1.8) mHz by a worst-case scenario. Unlike the first-order Zeeman shift, the quadratic Zeeman effect causes a net shift in the average frequency of a symmetric pair and it can be calculated using second-order perturbation theory. For the background bias field of 2257.4(5.0) nT used in our system, the net frequency shift is calculated to be 73.48(33) µHz. Similarly, the magnetic field associated with BBR can also produces a quadratic Zeeman shift. The BBR field for the effective temperature of 293.74(0.41) K obtained above gives $B_{rms} = 2717.0(3.8)$ nT and the related quadratic Zeeman shift of 106.44(31) µHz[27].

Electromagnetic fields of the lasers incident on the ion during the interrogations can cause ac Stark shift of the clock transition. For 397 nm laser, two shutters are employed to switch off the laser beam. With the first shutter is kept closed during the interrogations, the frequency difference between the states of the second shutter on and off is measured to be 1.67(16) Hz. Therefore, with an attenuation of >50 dB for the second shutter which switches off the 397 nm laser during the interrogations, the frequency shift is estimated to be 0.02(1) mHz by a worst-case scenario. For 854 nm laser, it's similar to the situation of 397 nm and the frequency shift is estimated with the same method to be -0.13(3) mHz. For the 866nm laser, only one shutter is used and the frequency difference between the states of the second shutter on and off is measured to be -29.89(33) Hz. Therefore, with an attenuation of >40 dB for the shutter which switches off the 866nm laser during the interrogations, the upper limit of frequency shift is estimated to be -2.99(4) mHz by a worst-case scenario. The ac Stark shift produced by the 729 nm laser is calculated by $\Delta\nu_{ac} = \kappa I$ where $\kappa$ is the coefficient of frequency shift as a function of laser intensity $I$ [29]. The average of parameter $\kappa$ for the selected three pairs of Zeeman components is 0.8(2)mHz/(W/m$^2$). In our current clock operation, the laser power incident on the ion is 7(3) nW focused to a spot waist of 137(20) μm, and the related frequency shift is calculated to be 0.19(10) mHz.

The frequency shift resulting from collisions with nonpolar gas molecules is estimated using a model based on phase changing Langevin collisions. As without a calibrated quadrupole mass analyzer assembled in our miniaturized vacuum chamber to measure the partial pressure of each gas components, an assumption is adopted that all the gas is composed of $H_2$. Therefore, the collisional frequency shift is calculated to be 1.7(1.7) mHz with a measured pressure of 3.0(3.0)×10$^{-8}$ Pa by a worst-case scenario[29].

TABLE. Ⅰ. Systematic shifts and uncertainties of the transportable clock.

| Source | Frequency shift | | Frequency uncertainty | |
| --- | --- | --- | --- | --- |
| | mHz | Fractional | mHz | Fractional |
| 2nd-order Doppler shift due to thermal motion | -8.6 | -2.09E-17 | 1.4 | 3.4E-18 |
| Stark shift due to thermal motion | 16.7 | 4.06E-17 | 2.8 | 6.8E-18 |
| Total excess micromotion shifts | -0.33 | -8E-19 | 0.58 | 1.4E-18 |
| Electric quadrupole shift | 0.0 | 0E-17 | 31 | 7.5E-17 |
| Stark shift due to blackbody radiation | 351.2 | 8.54E-16 | 5.3 | 1.3E-17 |
| Linear Zeeman shift due to bias field | 0.0 | 0E-18 | 1.8 | 4.4E-18 |
| 2nd-order Zeeman shift due to bias field | 0.073475 | 1.7875E-19 | 0.000077 | 1.9E-22 |
| 2nd-order Zeeman shift due to blackbody radiation | 0.10657 | 2.5927E-19 | 0.00034 | 8.3E-22 |
| Stark shift due to 397nm laser | 0.02 | 4.9E-20 | 0.01 | 2.4E-20 |
| Stark shift due to 854nm laser | -0.13 | -3.16E-19 | 0.03 | 7.3E-20 |
| Stark shift due to 866nm laser | -2.99 | -7.274E-18 | 0.04 | 9.7E-20 |
| Stark shift due to 729nm laser | 0.19 | 4.6E-19 | 0.10 | 2.4E-19 |
| Collisional shift | 1.7 | 4.1E-18 | 1.7 | 4.1E-18 |
| Total shifts | 358 | 8.71E-16 | 32 | 7.7E-17 |

Table Ⅰ summarizes frequency shifts and the related uncertainty contributions of our transportable $^{40}$Ca$^+$ single-ion clock. Considering all of them, we get a total shift of 358(32) mHz, which is 7.7×10$^{-17}$ of a fractional error. Fig. 5 shows an Allan deviation as a function of averaging time using self-comparison for a pulse time of 20 ms and average dead time of 30 ms. The observed long-term stability of $2.3\times10^{-14}/\sqrt{\tau}$ is in reasonable agreement with the calculated quantum projection noise (QPN) limit of $2.0\times10^{-14}/\sqrt{\tau}$ [34]. But the short-term stability of the clock is almost an order of magnitude worse than the 729 nm clock laser. The clock laser system has shown a good performance in our underground lab, but the evaluation of clock running is carried out after the transportable clock moved to another lab on the first floor and near the road. This lab is always with discontinuous interference of vibration to the clock laser, so a high value of servo gain is employed in the clock running to achieve a tightly locking. Then a large step of correction of frequency will lead to the degradation of short-term stability of clock laser.

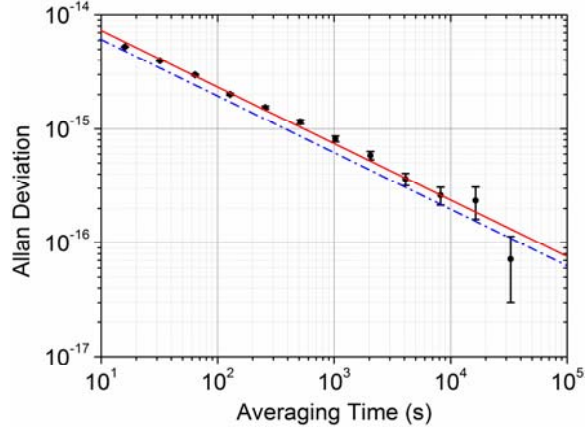

FIG. 5. Allan deviation of the transportable clock as a function of averaging time. The data shown as dots is obtained using self-comparison measurements with a probe pulse time 20 ms and red solid line represents a stability of $2.3\times10^{-14}/\sqrt{\tau}$. The QPN limit of $2.0\times10^{-14}/\sqrt{\tau}$ is shown by the blue dash-dot line.

## 4. Conclusion

In summary, we have shown a transportable optical clock refer to the $4s^2S_{1/2}$-$3d^2D_{5/2}$ electric quadrupole transition at 729 nm of $^{40}Ca^+$ single ion. The whole equipment of this transportable clock, consisting of a compact single ion unit and a compact laser unit, is constructed within a volume of 0.54 m$^3$ excluding the electronics. This kind of modular design, allowing for independent testing and maintenance, ensures better stability and reliability essential for long-term operation of clock when it moves to another place. The systematic fractional uncertainty has been evaluated to be $7.7\times10^{-17}$, and the stability fits to be $2.3\times10^{-14}/\sqrt{\tau}$ by clock self-comparison with a probe pulse time 20 ms. This performance, exceeding the state-of-art Cs primary standard by which the accuracy of SI second is currently realized, makes it a strong candidate for improving worldwide international atomic time (TAI) and widely using as a potential probe for sensitive phenomena. Of course, there's a great potential for improving the performance of our transportable clock. The maximum probe time and thus the stability of clock is limited by the coherence time of the clock laser and the ion. Therefore, a better controlling of the frequency noise of clock laser and motional state of single ion in Paul trap is one of the main direction in our research in the near future. In addition, we will also put in a lot of energy to research on the critical issues such as integration, reliability and automatic operation as practical transportability calls for compactness, robustness and intelligence on a transportable clock.

We acknowledge Yao. Huang, Hua. Guan, Liang. Chen and Fei. Zhou for their fruitful discussions. This work is supported by the National High Technology Research and Development Program of China (863 Program, Grant No. 2012AA120701) and the National Natural Science Foundation of China (Grant No. 11174326).

**References**

[1] Ludlow A D, Boyd M M, Ye J, et al. Optical atomic clocks[J]. Reviews of Modern Physics, 2015, 87(2): 637-701.

[2] Huntemann N, Sanner C, Lipphardt B, et al. Single-Ion Atomic Clock with $3\times10^{-18}$ Systematic Uncertainty[J]. Physical Review Letters, 2016, 116(6): 063001.

[3] Chou C W, Hume D B, Koelemeij J C J, et al. Frequency Comparison of Two High-Accuracy $Al^+$ Optical Clocks[J]. Physical Review Letters, 2010, 104(7): 070802.

[4] Nicholson T L, Campbell S L, Hutson R B, et al. Systematic evaluation of an atomic clock at $2 \times 10^{-18}$ total uncertainty[J]. Nature Communications, 2015, 6: 8.

[5] Ushijima I, Takamoto M, Das M, et al. Cryogenic optical lattice clocks[J]. Nat Photonics, 2015, 9(3): 185-189.

[6] Godun R M, Nisbet-Jones P B R, Jones J M, et al. Frequency Ratio of Two Optical Clock Transitions in $^{171}Yb^+$ and Constraints on the Time Variation of Fundamental Constants[J]. Physical Review Letters, 2014, 113(21): 210801.

[7] Huntemann N, Lipphardt B, Tamm C, et al. Improved Limit on a Temporal Variation of $m_p/m_e$ from Comparisons of $Yb^+$ and Cs Atomic Clocks[J]. Physical Review Letters, 2014, 113(21): 210802.

[8] Joao S M, Martins C, Mota I, et al. On the stability of fundamental couplings in the Galaxy[J]. Physics Letters B, 2015, 749: 389-392.

[9] Vutha A. Optical frequency standards for gravitational wave detection using satellite Doppler velocimetry[J]. New Journal of Physics, 2015, 17: 6.

[10] Chou C W, Hume D B, Rosenband T, et al. Optical Clocks and Relativity[J]. Science, 2010, 329(5999): 1630-1633.

[11] Bondarescu R, Scharer A, Lundgren A, et al. Ground-based optical atomic clocks as a tool to monitor vertical surface motion[J]. Geophysical Journal International, 2015, 202(3): 1770-1774.

[12] Fortier T M, Kirchner M S, Quinlan F, et al. Generation of ultrastable microwaves via optical frequency division[J]. Nature Photonics, 2011, 5(7): 425-429.

[13] Petit G, Arias F, Panfilo G. International atomic time: Status and future challenges[J]. Comptes Rendus Physique, 2015, 16(5): 480-488.

[14] Hachisu H, Fujieda M, Nagano S, et al. Direct comparison of optical lattice clocks with an intercontinental baseline of 9000 km[J]. Optics Letters, 2014, 39(14): 4072-4075.

[15] Predehl K, Grosche G, Raupach S, et al. A 920-kilometer optical fiber link for frequency metrology at the 19th decimal place[J]. Science, 2012, 336(6080): 441-444.

[16] Schiller S, Gorlitz A, Nevsky A, et al. The Space Optical Clocks Project: Development of high-performance transportable and breadboard optical clocks and advanced subsystems[J]. 2012 European Frequency and Time Forum (Eftf), 2012: 412-418.

[17] Poli N, Schioppo M, Vogt S, et al. A transportable strontium optical lattice clock[J]. Applied Physics B, 2014, 117(4): 1107-1116.

[18] Riehle F. Towards a redefinition of the second based on optical atomic clocks[J]. Comptes Rendus Physique, 2015, 16(5): 506-515.

[19] Bongs K, Singh Y, Smith L, et al. Development of a strontium optical lattice clock for the SOC mission on the ISS[J]. Comptes Rendus Physique, 2015, 16(5): 553-564.

[20] Hill I R, Hobson R, Bowden W, et al. A low maintenance Sr optical lattice clock[J]. arXiv:1602.05810, 2016.

[21] Huang Y, Guan H, Liu P, et al. Frequency Comparison of Two $^{40}Ca^+$ Optical Clocks with an Uncertainty at the $10^{-17}$ Level[J]. Physical Review Letters, 2016, 116(1): 013001.

[22] Cao Jian, T X, Cui Kai-Feng, Shang Jun-Juan, Shu Hua-Lin, Huang Xue-Ren. Simulation and Optimization of Miniature Ring-Endcap Ion Traps[J]. Chin. Phys. Lett., 2014, 31(04): 43701-.

[23] Berkeland D J, Miller J D, Bergquist J C, et al. Minimization of ion micromotion in a Paul trap[J]. Journal of Applied Physics, 1998, 83(10): 5025-5033.

[24] Safronova M S, Safronova U I. Blackbody radiation shift, multipole polarizabilities, oscillator strengths, lifetimes, hyperfine constants, and excitation energies in $Ca^+$[J]. Physical Review A, 2011, 83(1): 012503.

[25] Drever R W P, Hall J L, Kowalski F, et al. Laser phase and frequency stabilization using an optical resonator[J]. Applied Physics B, 1983, 31(2): 97-105.

[26] Bernard J E, Madej A A, Marmet L, et al. Cs-based frequency measurement of a single, trapped ton transition in the visible region of the spectrum[J]. Physical Review Letters, 1999, 82(16): 3228-3231.

[27] Dubé P, Madej A A, Zhou Z, et al. Evaluation of systematic shifts of the $^{88}$Sr$^+$ single-ion optical frequency standard at the $10^{-17}$ level[J]. Physical Review A, 2013, 87(2): 023806.

[28] Urabe S, Watanabe M, Imajo H, et al. Observation of Doppler sidebands of a laser-cooled Ca+ ion by using a low-temperature-operated laser diode[J]. Applied Physics B-Lasers and Optics, 1998, 67(2): 223-227.

[29] Madej A A, Bernard J E, Dubé P, et al. Absolute frequency of the $^{88}$Sr$^+$ 5s $^2$S$_{1/2}$ - 4d $^2$D$_{5/2}$ reference transition at 445THz and evaluation of systematic shifts[J]. Physical Review A, 2004, 70(1): 012507.

[30] Itano W M. External-field shifts of the $^{199}$Hg$^+$ optical frequency standard[J]. Journal of Research of the National Institute of Standards and Technology, 2000, 105(6): 829-837.

[31] Dolezal M, Balling P, Nisbet-Jones P B R, et al. Analysis of thermal radiation in ion traps for optical frequency standards[J]. Metrologia, 2015, 52(6): 842-856.

[32] Barwood G P, Huang G, Klein H A, et al. Agreement between two $^{88}$Sr$^+$ optical clocks to 4 parts in $10^{17}$[J]. Physical Review A, 2014, 89(5): 050501.

[33] Itano W M, Lewis L L, Wineland D J. Shift of $^2$S$_{1/2}$ hyperfine splittings due to blackbody radiation[J]. Physical Review A, 1982, 25(2): 1233-1235.

[34] Itano W, Bergquist J, Bollinger J, et al. Quantum projection noise: Population fluctuations in two-level systems[J]. Physical Review A, 1993, 47(5): 3554.